\documentstyle[12pt,twoside,fleqn,espcrc1,epsf]{article}

\newcommand{\AmS}{{\protect\the\textfont2
  A\kern-.1667em\lower.5ex\hbox{M}\kern-.125emS}}

\title{Spin Physics with the PHENIX Detector System}

\author{N. Saito\address{
        The Institute of Physical and Chemical Research, (RIKEN),
        Wako, Saitama, 351-01, Japan}
        for the {\sl PHENIX Collaboration\thanks{For the complete list
        of authors, please refer to Ref.~[1]. 
        Visit {\tt http://www.rhic.bnl.gov/phenix} for the most
        current PHENIX information.}}}

\begin{document}
\maketitle

\begin{abstract}
The PHENIX experiment at RHIC has extended its scope 
to cover spin physics using polarized proton beams.
The major goals of the spin physics at RHIC are 
elucidation of the spin structure 
of the nucleon and precision tests of the symmetries.
Sensitivities of the spin physics measurements with 
the PHENIX detector system are reviewed.
\end{abstract}

\section{PHENIX and RHIC Spin Physics} 
The Relativistic Heavy Ion Collider (RHIC) is under construction at
Brookhaven National Laboratory. 
In addition to its primary purpose, the search for quark-gluon plasma, 
a proposal to explore spin physics at RHIC~\cite{PROP} has been approved. 
The major goals of the spin physics program at RHIC are:
\begin{itemize} 
   \item elucidation of the spin structure of the nucleon, and 
   \item precision tests of symmetries. 
\end{itemize} 
RHIC offers a unique opportunity for those studies 
because of its capability of
accelerating polarized proton beams up to 
$\sqrt{s}$= 500~GeV at the luminosity of
2$\times$10$^{32}$cm$^{-2}$sec$^{-1}$ or more with 
large polarizations of $\sim$70\%.  
Obviously we will reach the high-energy frontier for polarized 
proton-proton collisions at RHIC. 

The PHENIX detector is one of the two large detectors at 
RHIC~\cite{PHENIX}.
Its basic design concept is to detect photons, leptons, and hadrons
with high momentum resolution and strong particle identification.  
It consists of  two spectrometers covering the central 
rapidity region (Central Arms), which include an electromagnetic
(EM) calorimeter with 
fine segmentation ($\Delta \eta \sim \Delta \phi \sim 0.01$), 
and two endcap muon spectrometers (Muon Arms). 
Since hadron reactions with photonic or leptonic
final states such as prompt photon production and lepton production
from weak boson decays play major roles in spin physics program, 
PHENIX is very well suited to spin physics at RHIC.

The studies are done by measuring the 
spin asymmetries in the cross sections for 
various reactions. 
By use of the spin rotators located upstream and downstream of 
PHENIX experimental hall, any combination of beam polarizations 
is possible. 
Thus we can measure the {\it helicity} dependent cross sections
$\sigma(++)$, $\sigma(--)$, $\sigma(+-)$, and $\sigma(-+)$ separately, where
$+$ and $-$ represent positive and negative helicity states of the 
beam particles
as well as {\it transversity} dependent cross sections,
$\sigma(\uparrow\uparrow)$, $\sigma(\downarrow\downarrow)$, 
$\sigma(\uparrow\downarrow)$, $\sigma(\downarrow\uparrow)$, 
where $\uparrow$ and $\downarrow$ represent transverse polarization 
of the beam particles. Among these asymmetries, we will discuss only two 
asymmetries in this presentation,
a double longitudinal-spin asymmetry, 
${\cal A}_{LL}$, and a single longitudinal-spin asymmetry, ${\cal A}_L$;
\begin{equation} 
{\cal A}_{LL}=\frac{\sigma(++) + \sigma(--) - \sigma(+-) - \sigma(-+)}
{\sigma(++) + \sigma(--) + \sigma(+-) + \sigma(-+)},~~~{\rm and}~~~
{\cal A}_{L} = \frac{\sigma(-) - \sigma(+) }{\sigma(-) + \sigma(+) }.
\end{equation} 
The quantity ${\cal A}_{LL}$  is often used to extract helicity 
dependent structure functions;
${\cal A}_L$
extracts parity violation effects in the reaction.
In the following section, the sensitivity of our measurements is
calculated assuming integrated luminosity of 320~pb$^{-1}$ and 
800~pb$^{-1}$ for $\sqrt{s}=200$~GeV and 500~GeV, respectively,
which corresponds to 10~weeks of running with 70\% machine 
efficiency. 

\section{Spin Structure of the Nucleon} 
The results of polarized muon scattering off a polarized
proton target reported by the EMC collaboration
have stimulated both experimental and
theoretical works to elucidate the spin structure of the proton.
The fraction of the proton spin carried by quarks, 
$\Delta \Sigma$=0.12$\pm$0.09$\pm$0.14, was
amazingly small comparing to the canonical expectation
0.60$\pm$0.12~\cite{JAFFE92}.
Post-EMC experiments, which have provided data with much better
statistics including deuterium and $^3$He targets, 
have confirmed the EMC results.
A recent global analysis gives $\Delta \Sigma \approx 0.3$,
and thus still shows a significant deficit. 

PHENIX will measure the gluon polarization 
$\Delta G(x)$ and anti-quark polarization 
$\Delta \bar{q}_{i}(x)$ with flavor $i$ identified 
not only to search for 
the origin of the deficit but also to check the 
validity of assumptions in the 
analysis to obtain $\Delta \Sigma$, e.g. SU(3)$_{\rm flavor}$.  
These measurements will be described in the 
following subsections.

\subsection{Gluon polarization} 
One of the most reliable channels to measure the gluon polarization 
is high $p_T$ prompt photon production. The production is dominated 
by the gluon Compton process, followed in significance by the annihilation process. 
By neglecting the contribution from annihilation 
channel, (which is justified in several 
predictions~\cite{DG}),
the asymmetry ${\cal A}_{LL}$ can 
be written at the leading order (LO) as a function of photon $p_T$, 
\begin{equation} 
{\cal A}_{LL}(p_{T})\approx \frac{\Delta G(x_{T})}{G(x_{T})} 
A_{1}^{p}(x_{T})
a_{LL}^{\theta^{*} = \frac{\pi}{2}} (gq\rightarrow \gamma q), 
~A_{1}^{p}(x_{T})=\frac{g_{1}^{p}(x_{T})}{f_{1}^{p}(x_{T})}=
\frac{\Sigma_{i} e_{i}^{2} \Delta q_{i}(x_{T})}
{\Sigma_{i} e_{i}^{2} q_{i}(x_{T})} .
\label{E:AsymDG}
\end{equation} 
Here $x_{T}=2 p_{T}/\sqrt{s}$, $\theta^{*}$ stands for the  
scattering angle of partons in their CMS, and $a_{LL}$ 
represents the double longitudinal spin asymmetry for the 
parton cross sections.  It should be noted that 
the PHENIX acceptance ($|\eta|\le 0.35$) strongly selects the samples 
in symmetric quark-gluon scattering at  
$\theta^* \sim \frac{\pi}{2}$ and this 
selection allows great simplification of the expression in  
Eq.~(\ref{E:AsymDG})~\cite{DG}. 
Since $a_{LL}(gq \rightarrow \gamma q)$ is 
calculated in QCD and $A_{1}^{p}(x)$  
has been measured in lepton scattering experiments,  
$\Delta G(x)$ can be extracted from the measured ${\cal A}_{LL}$. 

To overcome experimental difficulties due to the huge background 
from hadron decays, {\sc PHENIX's} finely segmented EM calorimeter plays
a crucial role in avoiding the fake prompt photon signal that results
from the merging of  two photons from a high-$p_T$ $\pi^0$. 
Since the PHENIX calorimeter is as fine as 
$\Delta \eta \sim \Delta \phi \sim 0.01$, the prompt photon can be identified 
up to 30~GeV/$c$ or more without serious background. 

The yield for the assumed integrated luminosities has been calculated 
using 
{\sc Pythia} for the PHENIX acceptance
and listed in Table~\ref{T:DG_SENSITIV}
for both $\sqrt{s}=$200~GeV and 500~GeV. 
In addition, the sensitivity of the
measurement of $\Delta G(x)$ has been evaluated
using $a_{LL}$ and the measured $A_{1}^{p}(x)$. 
The listed errors are statistical
only. We have identified the origin of the systematic errors
and have begun studies to minimize them. In addition, 
studies of $\Delta G(x)$ measurements with other channels such as 
$\pi^{0}$, open charm/beauty, and 
heavy quarkonium production are in progress. 

\vspace{-2em}
\begin{table}[hbt]
\begin{center}
\caption{Sensitivity summary for the measurements of gluon polarization
via prompt $\gamma$ production.}
\begin{tabular}{c|lll|lll}
\hline 
          & \multicolumn{3}{c|}{$\sqrt{s}=200$~GeV} 
          & \multicolumn{3}{c}{$\sqrt{s}=500$~GeV} \\
\hline
photon $p_T$ & yield & \multicolumn{2}{c|}{errors on} 
             & yield & \multicolumn{2}{c}{errors on} \\ 
(GeV/$c$)    &       & ${\cal A}_{LL}$ & $\Delta G/G$ 
             &       & ${\cal A}_{LL}$ & $\Delta G/G$  \\
\hline 
10---15 & $1.0\times 10^5$ & 0.006 & 0.05 & $9.0\times 10^5$ & 0.002 & 0.05 \\
15---20 & $1.3\times 10^4$ & 0.017 & 0.09 & $1.8\times 10^5$ & 0.005 & 0.06 \\
20---25 & $2.7\times 10^3$ & 0.038 & 0.17 & $5.3\times 10^4$ & 0.009 & 0.08 \\
25---30 & $5.9\times 10^2$ & 0.080 & 0.31 & $1.9\times 10^4$ & 0.015 & 0.12 \\
\hline
\end{tabular}
\label{T:DG_SENSITIV}
\end{center} 
\end{table}

\vspace{-3em}
\subsection{Anti-quark polarization}
The polarized-DIS experiments are sensitive to neither differences 
between anti-quarks and quarks nor their flavors, since the
photon couples to the square of their electric charge. Therefore 
the measurement of anti-quark polarization and the flavor 
decomposition will improve the knowledge on the spin of the 
nucleon significantly. 

The parity violating asymmetry ${\cal A}_{L}$ for $W$ production is 
presented at LO as
\begin{equation}
{\cal A}_{L}(W^{+}) = \frac{\Delta u(x_{1}) \bar{d}(x_{2}) -
                          \Delta \bar{d}(x_{1}) u(x_{2})}
                          {u(x_{1}) \bar{d}(x_{2}) + 
                          \bar{d}(x_{1}) u(x_{2})}
\end{equation}
For $W^{-}$ production, the $u$ and $d$ should be exchanged in this expression. 
The asymmetry is just the linear combination of the
quark and anti-quark polarizations. Furthermore, the flavor in the reaction 
is almost fixed. Thus flavor decomposition is possible. The asymmetry
converges to $\frac{\Delta u(x)}{u(x)}$ at the limit of $x_{1}\gg x_{2}$
and to $\frac{\Delta \bar{d}(x)}{\bar{d}(x)}$ at the limit of
$x_{2}\gg x_{1}$.

\begin{figure}[hbt]
\epsfxsize=4.0in
\begin{center}
\mbox{\epsfbox{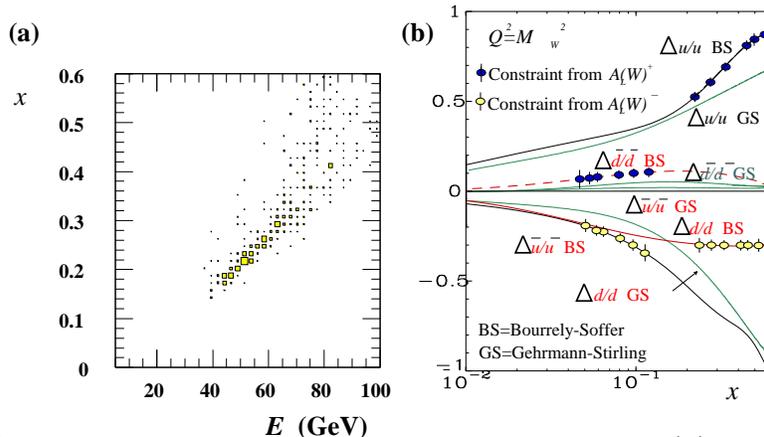}}
\vspace{-3em}
\caption{(a) Correlation between $x$ and muon energy $E^{\mu}$ and 
(b) models on polarized parton distributions and projected
statistical error. }
\label{F:POLPDF}
\end{center}
\end{figure}

$W^{\pm}$ production can be identified by the 
detection of muon with $p_T \ge$20~GeV/$c$. 
With the assumed $\int {\cal L} dt$=800~pb$^{-1}$,
we expect about 5000 events for each of $W^+$ and $W^-$. 
Furthermore the $x$-range can be selected using 
the muon momentum as shown in Figure~\ref{F:POLPDF}(a). 

Using the muon sample which is divided into several 
energy bins (and thus $x$-bins), we have estimated the sensitivities of 
our measurements (with statistical errors only). The results are plotted 
on two model predictions for polarized structure 
functions~\cite{BS} in Figure~\ref{F:POLPDF}(b). Error bars 
indicate the expected 
statistical errors. Our measurement will be quite sensitive to both 
quark and anti-quark helicity distributions, although further studies are 
needed to minimize the background contributions.  

\section{Symmetry tests}
The nature of parity non-conservation itself can be directly probed using the
polarized beams at RHIC, with ${\cal A}_{L}$ the measure of violation. 
Taxil and Virey studied various possibility to find 
new physics at RHIC through the measurements of ${\cal A}_L$
using polarized proton beams~\cite{TAXIL}. 
While their predictions are only for single-jet production, which is 
not detectable with PHENIX, we expect that such asymmetries, however, should 
persist
with 
inclusive or leading particle production.  Sensitivity
studies for PHENIX are underway.

\vspace{-3em}

\vspace{-3em}
\section{Summary} 
The PHENIX physics scope has been extended to include
the spin physics program. It
will provide measurements of $\Delta G(x)$ and $\Delta \bar{q}_{i}(x)$
that greatly reduce the uncertainties in the knowledge of 
the spin structure of the nucleon. Precision tests of symmetries in 
the standard model is foreseen. 
The first collision of the polarized proton beams is expected 
to start in the year of 2000. 

\section*{Acknowledgment} 
        We thank the technical staffs of the participating institutions 
for their vital contributions.  This detector construction project is
supported by the Department of Energy (U.S.A.), STA and Monbu-sho (Japan), 
RAS, RMAE, and RMS (Russia), BMBF (Germany), FRN and the Knut \& Alice 
Wallenberg Foundation (Sweden), and MIST and NSERC (Canada).


\begin{thebibliography}{9}
\bibitem{PHENIX} D.P.~Morrison for the PHENIX Collaboration,
these proceedings. 
\bibitem{PROP}Proposal on spin physics using the RHIC (R5),
Aug. 1992; updated Sep. 1993.
\bibitem{JAFFE92}R.L. Jaffe, Nucl. Phys. {\bf A547} (1992) 17c.   
\bibitem{DG}L.E. Gordon, Nucl. Phys. {\bf B501} (1997) 197. 
\bibitem{BS}C.~Bourrely and J.~Soffer Nucl. Phys. {\bf B445} (1995) 341.
\bibitem{TAXIL}P. Taxil and J.M. Virey Phys. Lett. {\bf B364} (1995)
181, {\bf B383} (1996) 355.


\end{thebibliography}
\end{document}